\documentclass[aps,pra,superscriptaddress,preprint,amsmath,amssymb,showpacs]{revtex4-1}
\usepackage[english]{babel}
\usepackage[cp866]{inputenc}
\usepackage{amssymb,bm}
\usepackage{graphicx}
\usepackage{amsmath}
\usepackage{color}
\usepackage[colorlinks=true,citecolor=blue,urlcolor=blue]{hyperref}
\begin{document}

\title{Parity violation in proton-deuteron scattering.}
\author{A.I. Milstein}
\affiliation{Budker Institute of Nuclear Physics of SB RAS, 630090 Novosibirsk, Russia}
\affiliation{Novosibirsk State University, 630090 Novosibirsk, Russia}
\author{N.N. Nikolaev}
\affiliation{L.D. Landau Institute for Theoretical Physics RAS, 142432 Chernogolovka, Russia}
\author{S.G. Salnikov}
\affiliation{Budker Institute of Nuclear Physics of SB RAS, 630090 Novosibirsk, Russia}
\affiliation{Novosibirsk State University, 630090 Novosibirsk, Russia}
\date{\today}

\begin{abstract}
	The effects of parity violation in the interaction of relativistic polarized protons and deuterons are discussed. Within Glauber's approach, estimates are obtained for P-odd asymmetries in the total and elastic scattering cross sections, in the deuteron dissociation cross section, and in the inelastic cross section with meson production in a final state. It is shown that, from the point of view of the magnitude of the P-odd effects, the interaction of polarized deuterons with unpolarized protons has an advantage over the interaction of polarized protons with unpolarized deuterons. A significant P-odd asymmetry was found in the dissociation channel of the polarized deuteron.
	\end{abstract}

\maketitle

{\bf Introduction.}
The interference of the amplitudes of strong and weak interactions leads to parity violation in nuclear and hadronic processes \cite{Abov,Lobashov}. The observed effects in nuclear processes and in the scattering of low-energy protons and neutrons are usually described by phenomenological meson-baryon interactions (see the review \cite{GHH2017}). Despite the extensive theoretical \cite{BHK1974,HK1975,KF1975, FK1976,DDH1980,FS1981,BT1981,O1981,NP1982,GP1983} and experimental \cite{P1974,N1978,B1980,L1984,Y1986,E1991,B2003} literature, the issue of parity violation in hadronic processes at high energies remains still open. Significant progress in understanding this effect can be expected from polarization experiments at the NICA collider \cite{Kekelidze2016vcp,Savin2014sva}. Possible experiments at NICA to search for parity violation in the interaction of longitudinally polarized protons or deuterons with an unpolarized target have been  discussed in Ref.~\cite{NICA}. Estimates of the P-odd asymmetry in nucleon-nucleon scattering in the NICA energy range have been reported given in our recent work \cite{MNS2020}. It is shown that, because of the structure of weak currents, the main contribution to the P-odd asymmetry in $pp$ scattering comes from radiative corrections due to the strong charge-exchange interactions. It has been also shown that, from the point of view of the magnitude of the observed effect, it is preferable to measure the P-odd asymmetry in elastic scattering, since the asymmetry in inelastic cross sections is strongly suppressed.

In this paper, we generalize the results of \cite{MNS2020} to P-odd asymmetries in proton-deuteron scattering at energies of the NICA collider. In contrast to nucleon-nucleon scattering, there is an important quasi-elastic scattering channel with the dissociation of a deuteron into a $pn$ continuum. Similar to the result of \cite{MNS2020}  on the enhancement of asymmetry in elastic scattering, a similar enhancement is found for the dissociation of a longitudinally polarized deuteron on an unpolarized proton. The P-odd asymmetry in the interaction of a polarized deuteron with an unpolarized proton turns out to be higher than in the interaction of a polarized proton with an unpolarized deuteron. This is important from the experimental point of view, since in the energy range of the NICA collider, the acceleration of polarized deuterons is free of spin resonances, which are numerous in the case of polarized protons. With regard to the separation of the P-odd asymmetry in the processes of elastic scattering and dissociation of accelerated deuterons, favorable from the point of view of the expected magnitude of the effect, we draw attention to the possibility of working with an internal jet hydrogen target with the detection of recoil protons \cite{Nikitin}.

{ \bf Nucleon-nucleon scattering.} 
A total amplitude 	$T(\bm q_\perp)$ of high energy elastic proton-nucleon scattering, where  $\bm q_\perp$ is the transverse momentum of the scattered proton, can be represented as \cite{MNS2020}
  \begin{eqnarray}\label{Mtot}
 &&T(\bm q_\perp)=T_s(\bm q_\perp)+T_W(\bm q_\perp)+ T_{int}(\bm q_\perp)\,,\nonumber\\
 &&  T_{int}(\bm q_\perp)=-\dfrac{i}{2} \int \dfrac{d^2q'_\perp}{(2\pi)^2}\,T_s(\bm q'_\perp) T_W(\bm q_\perp-\bm q'_\perp)\,. 
  \end{eqnarray}
 Here $T_s(\bm q_\perp)$ is the strong interaction scattering amplitude, $T_W(\bm q_\perp)$
is the weak interaction scattering  amplitude with account for the radiative corrections to the P-odd Hamiltonian due to the strong interactions, $T_{int}(\bm q_\perp)$ is the so-called absorption correction to a weak amplitude, it is not difficult to derive it in the eikonal approach. Taking into account the conservation of the $s$-channel helicity for the $pN$ scattering amplitudes (hereinafter $N=p,n$), one can use the standard  parameterization \cite{R2003} (the difference from the alternative  parameterizations \cite{SH2010,FO2013} is insignificant and is not being discussed): 
  \begin{align}\label{TS}
 T_s^{pN}(\bm q_\perp)=\delta_{\lambda_1\lambda_3}\delta_{\lambda_2\lambda_4}t_s^{pN}(\bm q_\perp)\ ,
 \quad t^{pN}_s(\bm q_\perp)=-(\epsilon_{pN}+i)\,\sigma_{s,\,tot}^{pN}\exp(-B_{pN}q_\perp^2/2)\, ,
 \end{align} 
 where  $\lambda_1$ and   $\lambda_2$ are the helicities of the initial particles, $\lambda_3$ and $\lambda_4$   are the corresponding helicities of the final particles ($\lambda_i = \pm 1$).  For  momentum transfers  inside the diffraction cone, the ratio $\epsilon_{pN}$ of the real and imaginary parts of the amplitude and the slope $B_ {pN}$ of the diffraction cone  can be considered constants. To a sufficient accuracy,  in the NICA energy range we can put $t^{pp}_s(\bm q_\perp)=t^{pn}_s(\bm q_\perp)\equiv t_s(\bm q_\perp)$, see Ref.~\cite{R2003}. In numerical estimates, we use
   \begin{align}\label{par}
  &\epsilon_{pN}=\epsilon=-0.5\,,\quad  \sigma_{s,\,tot}^{pN}=\sigma_{s,\,tot}  =50\,\mbox{mb}\,,\quad B_{pN}=B=9\,\mbox{GeV}^{-2}\,.
 \end{align}
In this case, the elastic scattering cross section is
\begin{eqnarray}\label{sigel}
\sigma_{s,\,el}^{pN}=\int | T_s^{pN}(\bm q_\perp)|^2\,\dfrac{d^2q_\perp}{16\pi^2}=\dfrac{(1+\epsilon^2)\sigma_{s,\,tot}^2}{16\pi B}=17.8\,\mbox{mb}\,.
\end{eqnarray}

According to \cite{MNS2020}, the amplitudes due to the weak interaction, $T_W^{pN}(\bm q_\perp)$, have different dependences on the momentum transfer and helicities:
  \begin{align}
 &T_W^{pp}(\bm q_\perp)=\lambda_1\delta_{\lambda_1\lambda_2}\delta_{\lambda_1\lambda_3} \delta_{\lambda_1\lambda_4}t_W^{pp}(\bm q_\perp)\,,\quad  T_W^{pn}(\bm q_\perp)=\lambda_1\delta_{\lambda_1\lambda_3}\delta_{\lambda_2\lambda_4}t_W^{pn}(\bm q_\perp)\,,\nonumber\\
 &   t_W^{pp}(\bm q_\perp)= c_{pp}\,R(\bm q_\perp)\,,\quad t_W^{pn}(\bm q_\perp)=   c_{pn}\,F^2(\bm q_\perp)\,, \nonumber\\
 &F(\bm q_\perp)=\dfrac{\Lambda^4 }{(\Lambda^2+q_\perp^2)^2}\,,\quad R(\bm q_\perp)=\dfrac{4}{\pi}\int \dfrac{F^2(\bm k_\perp)\,d^2k_\perp}{(\bm k_\perp-\bm q_\perp)^2+m_\rho^2}\,,\nonumber\\
 &c_{pp}=5\,\mbox{nb}\,,\quad c_{pn}=-7.8\,\mbox{nb}\,,\quad \Lambda=1\,\mbox{GeV}\,,\quad m_\rho=770\,\mbox{MeV}\,.
 \end{align}
 Note that $c_{pp}$ and $c_{pn}$ have opposite signs.
 
 Using the optical theorem, $ \sigma_{tot}=-\mbox{Im}T(0)$, we found the corrections $\sigma_{W,\,tot}^{pp}$ and $\sigma_{W,\,tot}^{pn}$ to the total cross section for $pp$ and $pn$ scattering due to weak interaction:
  \begin{align}
 &\sigma_{W,\,tot}^{pp}=\lambda_1\delta_{\lambda_1\lambda_2}\delta_{\lambda_1\lambda_3} \delta_{\lambda_1\lambda_4}S_W^{pp}\,,\quad
 \sigma_{W,\,tot}^{pn}=\lambda_1\delta_{\lambda_1\lambda_3}\delta_{\lambda_2\lambda_4}  S_W^{pn}\,,\nonumber\\
 &S_W^{pp} =3.7\,\mbox{nb}\,,\quad S_W^{pn} =-2.47\,\mbox{nb}\,.                 
 \end{align} 
The ratio between $S_W^{pp}$ and $S_W^{pn}$ is determined not only by the ratio between $c_ {pp}$ and $c_ {pn}$, but also by different dependences  on $q_\perp$ of the amplitudes $t_W^{pp}$ and $t_W^{pn}$, see below.
With the simplified parametrization \eqref{TS}, the P-odd corrections $\sigma_{W,\,el}^{pN}$ to the elastic scattering $pN$ cross sections  coincide with the P-odd corrections to the corresponding total cross sections. The implied suppression of P-odd corrections to inelastic cross sections is, in essence, a general consequence of the unitarity condition in the approximation linear in the weak interaction. 

{ \bf Weak interaction effects in proton-deuteron scattering.}
 Our estimations are based on the Glauber's approach \cite{G1955,G1966,G1967}. A new channel of diffraction dissociation (quasi-elastic scattering) into a proton-neutron continuum without meson production, $pd\to p(pn)$, requires a special consideration. We  predict that  a large P-odd asymmetry   occurs  also   in quasi-elastic scattering.

The amplitude $T^{pd}_s$ of  elastic $pd$ scattering due to the strong interactions reads
\begin{align}
& T^{pd}_s(\bm q_\perp)=\delta_{\lambda_p\lambda_p'} \delta_{\lambda_d\lambda_d'}\, t^{pd}_s(\bm q_\perp)\,,\nonumber\\ & t^{pd}_s(\bm q_\perp)=
 \Big[t_s^{pp}(\bm q_\perp)+t_s^{pn}(\bm q_\perp)\Big]\,F_D\left(\dfrac{\bm q_\perp}{2} \right)\nonumber\\  
&-\frac{i}{2}\int\,\dfrac{d^2q_\perp'}{(2\pi)^2}\,t_s^{pp}\left(\dfrac{\bm q_\perp }{2}-\bm q'_\perp\right) t_s^{pn}\left(\dfrac{\bm q_\perp }{2}+\bm q'_\perp\right)  \,F_D\left(\bm q_\perp'\right)\,. 
\end{align} 
Here $\lambda_p$ and $\lambda_p'$ are the helicities of the initial and final state protons, $\lambda_d$ and $\lambda_d'$ are the   helicities of the initial and final state   deuterons.
A single scattering amplitude contains the deuteron form factor $F_D\left(\bm q_\perp/2\right)$, and the double scattering amplitude, which is small in the region of the diffraction cone of elastic $pd$ scattering, gives Glauber screening. The deuteron form factor can be estimated to a sufficient accuracy using the $S$-state wave function $\phi(\bm r)$:
$$F_D(\bm q)=\int\,d^3r\,|\phi(\bm r)|^2\,\exp(-i\bm q\cdot\bm r)\,.$$
The explicit form  of $F_D(\bm q)$     obtained in the model of  a square potential well is
\begin{align}
&F_D(q)=\dfrac{2 b}{(b-1)\,x}\Big[\arctan\left(\dfrac{x}{2}\right)-\dfrac{1}{2}\mbox{Si}\left(\dfrac{x}{b} \right)-\dfrac{1}{4}\mbox{Si}\left(\pi+\dfrac{x}{b} \right)
+\dfrac{1}{4}\mbox{Si}\left(\pi-\dfrac{x}{b} \right)\Big]\,,\nonumber\\
&\mbox{Si}(x)=\int_0^xdy\,\frac{\sin\,y}{y}\,,\quad b=2.5\,,\quad x=q/\kappa\,,\quad \kappa=45.7\,\mbox{MeV} \,.
\end{align} 
Numerically, this representation is in good agreement with those obtained in other models.

The total cross section of $pd$  scattering following from the optical theorem reads
\begin{align}\label{sigstot}
&\sigma_{s,\,tot}^{pd}=2\sigma_{s,\,tot}-\Delta \sigma_G=96\,\mbox{mb}\,,\nonumber\\
&\Delta \sigma_G=\frac{1}{2}(1-\epsilon^2)\sigma_{s,\,tot}^2\int\,\dfrac{d^2q_\perp}{(2\pi)^2}\,\exp(-B\,q_\perp^2)\,F_D\left(\bm q_\perp\right)= 4\,\mbox{mb}\,.
\end{align}
The Glauber screening correction, $\Delta \sigma_G$,   is small due to the large size of the deuteron, $\Delta \sigma_G\ll \sigma_{s,\,tot}^{pd}$. In view of the obvious dominance of the single scattering amplitude, the P-odd asymmetry in $ pd $ scattering will be similar to that   in elastic $ pN $ scattering. The integrated cross section of elastic $ pd $ scattering will be noticeably suppressed by the deuteron form factor. In the same approximation of a loose deuteron, the differential cross section for quasi-elastic $ pd $ scattering will be close to the sum of the differential cross sections for elastic $ pp $ and $ pn $ scattering. Correspondingly, we expect that the observation in \cite {MNS2020} on the enhancement of the P-odd asymmetry in elastic $pN$ scattering   will persist in both elastic and quasi-elastic $pd$ scattering. We omit the discussion of the deuteron charge-exchange process ($d\to pp$) having a negligible cross section.

The total contribution of the weak interaction to the amplitude of elastic scattering of a polarized proton by a polarized deuteron, $T^{pd}_{W}(\bm q_\perp)$, including all absorption corrections, equals 
 \begin{align}\label{TWlpld}
&T^{pd}_W(\bm q_\perp)=\delta_{\lambda_p\lambda_p'} \delta_{\lambda_d\lambda_d'}\,t^{pd}_W(\bm q_\perp)\,,\nonumber\\
&t^{pd}_W(\bm q_\perp)={\cal T}_{\lambda_p\lambda_d}(\bm q_\perp)\,F_D\left(\dfrac{\bm q_\perp}{2} \right)    -\frac{i}{2}\int\!\dfrac{d^2q'_\perp}{(2\pi)^2}\,t_s\left(  \bm q'_\perp\right){\cal T}_{\lambda_p\lambda_d}(\bm q_\perp-\bm q'_\perp)  \,F_D\left(\dfrac{\bm q_\perp}{2} \right)\nonumber\\
&-\frac{i}{2}\int\,\dfrac{d^2q_\perp'}{(2\pi)^2}\,t_s\left(\dfrac{\bm q_\perp }{2}-\bm q'_\perp\right){\cal T}_{\lambda_p\lambda_d}\left(\dfrac{\bm q_\perp }{2}+\bm q'_\perp\right) \,F_D\left(\bm q_\perp'\right)\nonumber\\
&-\frac{1}{4}\iint\!\dfrac{d^2q'_\perp}{(2\pi)^2}\, \dfrac{d^2q''_\perp}{(2\pi)^2}\,t_s\left(\dfrac{\bm q_\perp }{2}-\bm q'_\perp\right)\, t_s\left(\dfrac{\bm q_\perp }{2}-\bm q''_\perp\right){\cal T}_{\lambda_p\lambda_d}\left( \bm q'_\perp  +\bm q''_\perp\right) F_D \left(\bm q_\perp'  \right)\,,\nonumber\\
&{\cal T}_{\lambda_p\lambda_d}(\bm q_\perp)=\dfrac{1}{2}(\lambda_p+\lambda_d)\,t_W^{pp}(\bm q_\perp)+\lambda_p\,t_W^{pn}(\bm q_\perp)\,.
\end{align} 
The main P-odd contribution $\sigma_{W,\,el}^{pd}$ to the elastic scattering cross section is
\begin{align}
&\sigma_{W,\,el}^{pd}=\int\dfrac{d^2q_\perp}{8\pi^2}\,\mbox{Re}\Big[t^{pd\,*}_s(\bm q_\perp)t^{pd}_W(\bm q_\perp)\Big]\nonumber\\
&\simeq -
\dfrac{\epsilon\sigma_{s,\,tot}}{4\pi^2} \int{d^2q_\perp}\,\exp(-B\, q_\perp^2/2)\,{\cal T}_{\lambda_p\lambda_d}(\bm q_\perp)F_D^2\left(\bm q_\perp/2\right)\,.
\end{align}
One should use $\lambda_d=0$ for scattering of a polarized proton by an unpolarized deuteron and   $\lambda_p=0$  for  scattering of a polarized deuteron by an unpolarized proton.

Following the Franco-Glauber technique \cite{G1966,G1967}, one can readily   obtain a P-odd correction to the cross section for quasi-elastic $pd$ scattering. Omitting the details of calculations, we restrict ourselves to the statement that a sum of  elastic ($\sigma_{W,\,el}^{pd}$)  and quasi-elastic ($\sigma_{W,\,qel}^{pd}$) P-odd scattering cross sections    coincides with the correction $\sigma_{W,\,tot}^{pd}$ to the total cross section of $pd$ scattering, which can be determined from the amplitude \eqref{TWlpld} using the optical theorem:
\begin{align}
&\sigma_{W,\,tot}^{pd}= -
\dfrac{\epsilon\sigma_{s,\,tot}}{8\pi^2} \int{d^2q_\perp}\,\exp(-B\, q_\perp^2/2)\,{\cal T}_{\lambda_p\lambda_d}(\bm q_\perp)\,[1+F_D\left(\bm q_\perp\right)]\,.
\end{align}
As in the case of inelastic $pN$ scattering, the P-odd asymmetry in inelastic $pd$ scattering, in which mesons are produced, is suppressed.

\begin{figure}[h]
\centering
\includegraphics[width=0.65\linewidth]{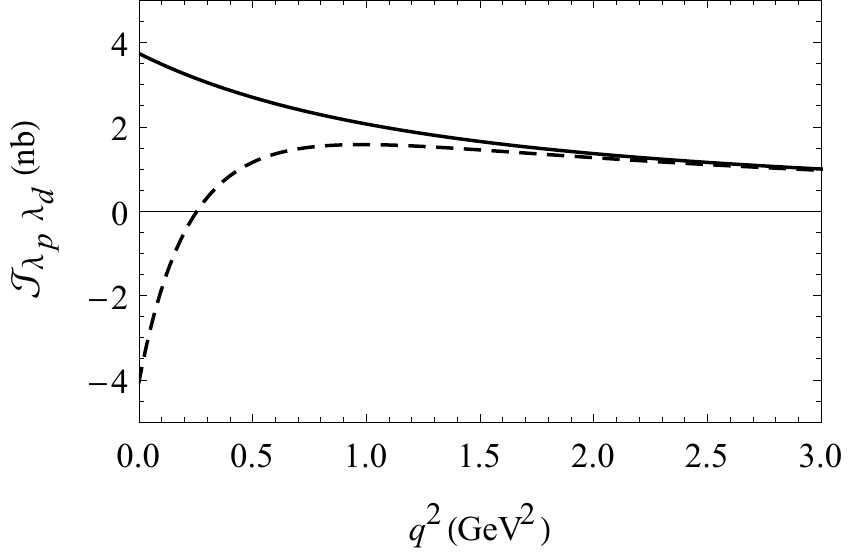}
\caption{Dependence of ${\cal T}_{0\,1}$  (solid line) and ${\cal T}_{1\,0}$ (dotted line) on $q^2\equiv \bm q^2_\perp$, Eq.~\eqref{TWlpld}.  }
\label{Tpd}
\end{figure}

Let us pass from a qualitative discussion to numerical estimates of the cross sections and corresponding asymmetries ${\cal A}=\sigma_W/\sigma_s\,$ in the scattering of a polarized deuteron with $\lambda_d=1$ by an unpolarized proton. Using the formulas obtained above, we find:
\begin{align}\label{delsigd}
&\sigma_{s,\,tot}^{pd} = 96\, \mbox{ mb},\, \quad \sigma_{W,\,tot}^{pd}=2.1\,\mbox{nb},\quad {\cal A}_{tot}^{pd}=2*10^{-8}\,, \nonumber\\
&\sigma_{s,\,el}^{pd} = 20\,\mbox{ mb},\quad \sigma_{W,\,el}^{pd}=0.7\,\mbox{nb}\,,\quad
{\cal A}_{el}^{pd}=3.5*10^{-8}\, , \nonumber\\
&\sigma_{s,\,qel}^{pd} = 22.4\mbox { mb}, \quad\sigma_{W,\,qel}^{pd}=1.4\,\mbox{nb}\,,\quad {\cal A}_{qel}^{pd}=6*10^{-8}\,.
\end{align}
For the interaction of a polarized proton with $\lambda_p=1$ and an unpolarized deuteron, we have
\begin{align}\label{delsigp}
& \sigma_{W,\,tot}^{pd}=-0.8\,\mbox{nb}\,,\quad {\cal A}_{tot}^{pd}=-0.9*10^{-8}\,,\nonumber\\
& \sigma_{W,\,el}^{pd}=-0.6\,\mbox{nb}\,,\quad {\cal A}_{el}^{pd}=-3*10^{-8}\,, \nonumber\\
& \sigma_{W,\,qel}^{pd}=-0.2\,\mbox{nb}\,,\quad {\cal A}_{qel}^{pd}=-10^{-8}\,.
\end{align}
The difference in signs and magnitudes of asymmetries is associated with a significant difference in the dependence of ${\cal T}_{\lambda_p\lambda_d}(q_\perp)$ on $q_\perp$ for polarized protons and for polarized deuterons, see Fig.~\ref{Tpd}.
Since the P-odd Hamiltonian of the weak $pp$ interaction is determined by the radiative
corrections due to strong interactions and the accuracy of calculating these radiative corrections is not high, the  behavior of ${\cal T}_{\lambda_p\lambda_d}(q_\perp)$ shown in Fig. \ref{Tpd}   is rather a qualitative one.

Two important conclusions follow from  estimates \eqref{delsigd} and \eqref{delsigp}. First, the magnitude of the expected P-odd asymmetry makes investigation of $pd$ scattering of polarized deuterons by unpolarized protons preferred one as compared to scattering of polarized  protons by unpolarized deuterons. It is also preferable from the point of view of controlling the polarization of particles in the accelerator, since deuterons have no spin resonances in the NICA energy range, while protons have numerous spin resonances. Second, because of the value of the expected asymmetry, it is preferable to separate elastic and quasi-elastic $ pd $ scattering. Here we will briefly comment on the ongoing analysis of the attractive possibilities of working with an internal jet hydrogen target, which is under consideration by the team of the RFBR grant No.~ 18-02-40092 MEGA, who   coauthored   \cite {NICA}.

When working with a jet target (see, e.g., Ref.~\cite{Nikitin}), to detect elastic $dp$ scattering it is sufficient to measure the  momentum transfer to the recoil proton, which is uniquely related to its scattering angle, $\theta = q_z/q_{\perp} = q_{\perp}/(2m_p)$. Dissociation of the relativistic deuteron with $\gamma \gg 1$ into a $np$ pair with the excitation energy $\epsilon^*$ gives an additional contribution to the longitudinal momentum of recoil protons, $\Delta q_z = \epsilon^*/\gamma$, which increases a scattering angle $\theta $. In this case, the distribution over the transverse momentum of recoil protons becomes  broader as well, in comparison with purely elastic scattering. This makes it possible to register quasi-elastic events with simultaneous discrimination of pion production events when $\epsilon^* > m_{\pi}$.

{\bf Conclusions.}
We have analyzed the effects of parity violation in the scattering of protons by deuterons at energies of the NICA collider. Using Glauber approach, estimates are obtained for the corrections due to the weak interaction to the total, elastic, inelastic and dissociation cross sections in $pd$ scattering, as well as the corresponding spin asymmetries, see \eqref{delsigd}   and \eqref{delsigp}. According to our results, experiments on the scattering of polarized deuterons by unpolarized protons are preferred. This circumstance is especially important, since the acceleration of relativistic polarized deuterons is simpler than the acceleration of polarized protons. The results obtained should be taken into account when planning experiments at the NICA collider.

We are grateful to I.A. Koop and Yu.M. Shatunov for
stimulating discussions.
 
This work was supported by the RFBR grant No.~18-02-40092 MEGA.

\end{document}